# Smart Motion Detection System Using Raspberry Pi

Venkat Margapuri

Department of Computer Science

Kansas State University

Manhattan, USA

**Abstract**

*This paper throws light on the security issues that modern day homes and businesses face and describes the implementation of a motion detection system using Raspberry Pi which could be an effective solution to address the security concerns. The goal of the solution is to provide an implementation that uses PIR motion sensors for motion detection and sends notifications to users via emails.*

Key Words: Motion Detection, Raspberry Pi, PIR Sensor, LED, Raspberry Pi Camera, UPPAAL Model Verification

## 1 Introduction

One of the most important concerns in the modern day world, be it for homes or businesses is security. An estimated 2,000,000 burglaries are reported each year in the United States out of which 66% are attributed to break-ins. An increased number of people in the work force limits the amount of time that people spend at home leaving home security vulnerable. In addition to break-ins, the rise of online shopping led to skyrocketed porch pirating in recent times. An estimated 25.9 million Americans reported porch pirating in 2017 which was up from 25.3 million in 2017. This paper focuses on addressing home and business security concerns by the providing a motion detection solution using Raspberry Pi and PIR Motion Detection Sensors.

The Raspberry Pi is a small-sized computer (almost the size of a credit-card) that has the ability to plug into a computer monitor or any other display and can be connected to a keyboard and mouse for operation. It has an operating system called Raspbian OS and can be a very handy system to run applications in programming languages like Scratch and Python. Although small, the Raspberry Pi is a system that approximates a desktop or a laptop in terms of functionality. The Raspberry Pi can also connect to the internet. Mundane activities performed on a desktop/ laptop such as browsing the internet, spreadsheet creation, word processing and gaming can all be performed on the Raspberry Pi.

Passive Infrared Sensors, typically referred to as PIR Sensors are tiny devices that aid in motion detection. The PIR Sensors are inexpensive, require low power, last long and high performance devices. For that reason, they could be embedded into larger devices such as cameras and other video capable devices to detect objects. These portable devices have the ability to detect levels of Infrared radiation which is the key principle behind motion detection described in the paper.

The further sections of the paper i.e. section 2 gives a high level description of the paper by describing the main components of the project and their working and section 3 discusses the two main use cases of the project namely image capturing and video recording. Section 4 describes the UPPAAL model used for the verification of the system. Section 5 describes the next steps for the system and section 6 concludes the paper.

## 2 System Overview

The PIR sensor emits an output anytime motion is detected within the range of its sensors. The Raspberry Pi executes a Python program that tracks input on one of the Pi's I/O pins. The indication of motion being detected is provided in multiple ways. They are:

- Indication on the monitor/ display with a message that reads, "Motion Detected".
- An LED is lit up on the breadboard.
- The Raspberry Pi camera captures an image and sends an email with the image as an attachment.
- The Raspberry Pi camera records video for a time period of 5 seconds and converts it from the native .h264 format to MP4 format for ease of consumption.

To display the message on the console, the input from the Pi's input pin is used by the Python program.

Lighting up of the LED is achieved by connecting the output pin of the PIR Sensors to the LED. The output pin of the PIR Sensor emits a signal when motion is detected which is then used by the LED to light up.

Capturing images, recording videos and sending emails out to users are achieved by the use of different tools compatible with Raspberry Pi and Python.

All of these mechanisms are discussed in the following sections of this paper.

**Hardware System Specifications**

- Raspberry Pi 3, Model B+, 1 GB RAM
- PIR Motion Sensors
- Raspberry Pi Camera
- Resistor to regulate voltage
- LED
- Breadboard
- Connector cables male to female and female to male
- Monitor to display output(could be a laptop)

**Software System Specifications**

- Linux based OS(Raspbian OS)
- Python and tools compatible with Python such as Config Parser and GPAC
- VNC Viewer
- Putty

### 2.1 Raspberry Pi Overview

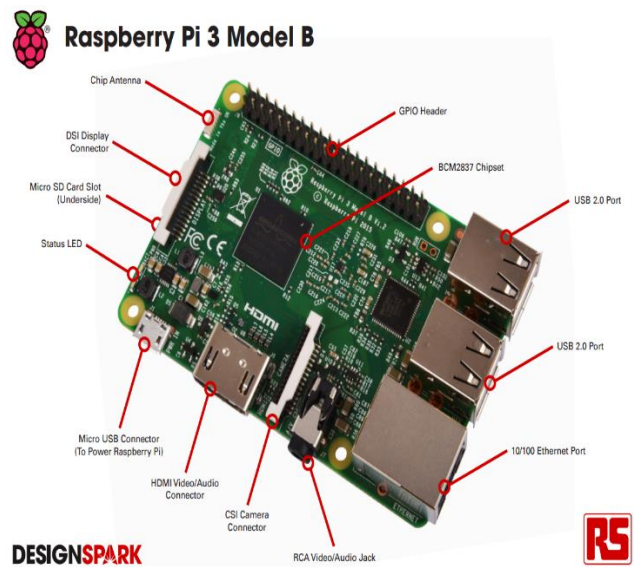

Figure 1: Raspberry Pi Device Description

The following are the most important components of Raspberry Pi 3:

- Ethernet port
- 4 USB 2.0 ports

- HDMI output
- GPIO (General Purpose Input/ Output)

The GPIO of the Raspberry Pi is the most critical component which make interactions with the breadboard (or any other external interaction) possible. A total of 40 GPIO pins are present on the Raspberry Pi. The voltages and the input/ output capabilities of the pins are as follows:

- **Voltages:** Two 5V pins and two 3V pins are present on the board, as well as a number of ground pins which are not configurable. The remaining pins are all general purpose 3V pins, meaning outputs are set to 3V and inputs are 3V tolerant.

Figure 2: GPIO Layout

- **Output:** A GPIO pin designated as an output pin can be set to high (3V) or low (0V).
- **Input:** A GPIO designated as an input pin can be as high (3V) or low (0V). Pull-up and pull-down resistors can be used for this purpose. Only pins GPIO2 and GPIO3 have fixed pull-up resistors, but the others have to be configured in software.

The GPIO pins and their numbers are not obvious at first glance. For this purpose, the pinout command could be used. The pinout command prints out the layout of the Raspberry Pi and helps understand the pins better.

The figure shows the usage of the 'pinout' command to understand the layout and the configuration of the Raspberry Pi.

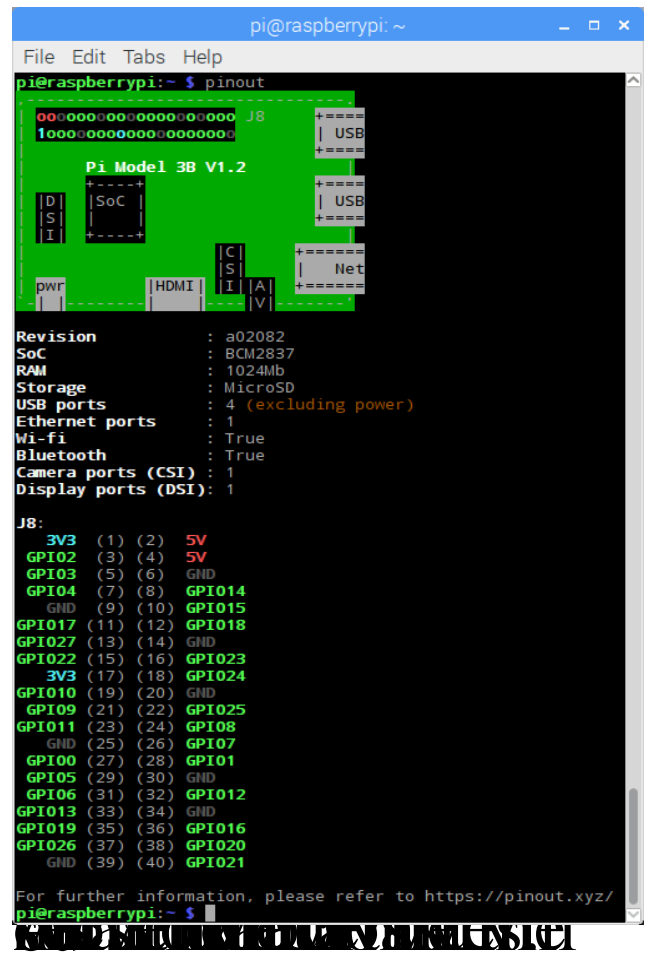

Figure 3: 'pinout' command

As you can see from the numbering of the pins in Figure 3, there are two different types of numberings assigned to the pins of the Raspberry Pi. They are typically referred to as modes and are important to understand to be able to write Python code to interact with the Raspberry Pi which will be demonstrated in the following sections.

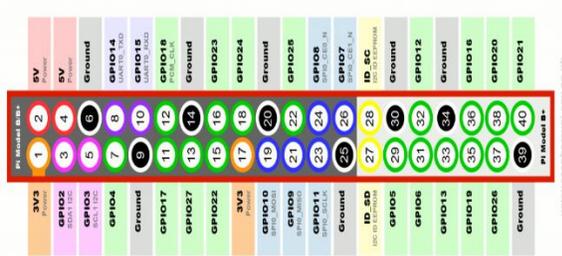

Figure 4: BOARD and BCM modes of the Raspberry Pi

There are two modes that are used to assign numbering to the pins on the Raspberry Pi, namely BOARD mode and BCM mode.

- **BOARD:** The BOARD mode specifies that the pins are being referred to by the number of the pin the the plug - i.e the numbers printed on the board (e.g. P1) and inside the Red rectangle as shown in the picture above.
- **BCM:** The BCM mode means that you are referring to the pins by the "Broadcom SOC channel" number, these are the numbers after "GPIO" outside the Red rectangle as shown in the diagram above.

## 2.2 PIR Motion Sensor Overview

PIR Motion Sensors are made of pyroelectric sensors and the sensors work on the principle of Pyroelectricity.

***Pyroelectricity is described as a property of materials where materials generate a certain amount of voltage when they are subject to temperature changes i.e. heated or cooled.***

Every entity, be it a living or a non-living entity emits a certain level of radiation and the amount of radiation emitted is directly proportional to the temperature of a particular entity (It is interesting to note that human beings emit radiation that is around 12 microns).

### 2.2.1 Components of PIR Motion Sensor

- **BIS0001 PIR Chip:** The 'BISS0001 PIR Chip' is a low-power CMOS technology chip.
- **Delay Time Adjust Knob:** The Delay Time Adjust knob determines the period of time for which the output of the PIR sensor remains high after detection of motion. It is adjustable and can be made to vary between 3 seconds and 5 minutes.

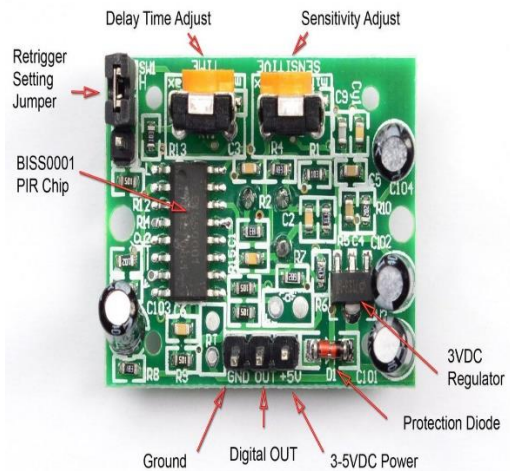

Figure 5: PIR Sensor Components

- **Sensitivity Adjust knob**: The sensitivity adjust knob determines the distance (range) of motion detection of the sensor. It can be made to vary between 3 meters and 7 meters. The view area of the sensors is typically 110 degrees.
*Note: The Delay Time Adjust and the Sensitivity Time Adjust decrease and increase upon clockwise and anti-clockwise rotations respectively.*

- **Digital OUT:** Digital pulse high (3V) when triggered (motion detected) digital low when idle (no motion detected).

### 2.2.2 Working of a PIR Motion Sensor

The PIR Motion Sensor has two slots, each slot made of a material that is sensitive to Infrared radiation and a lens. Both the slots can detect motion within the sensitivity range of the sensors. In their idle state where the sensors are exposed to ambient conditions, both the slots detect the same amount of IR radiation.

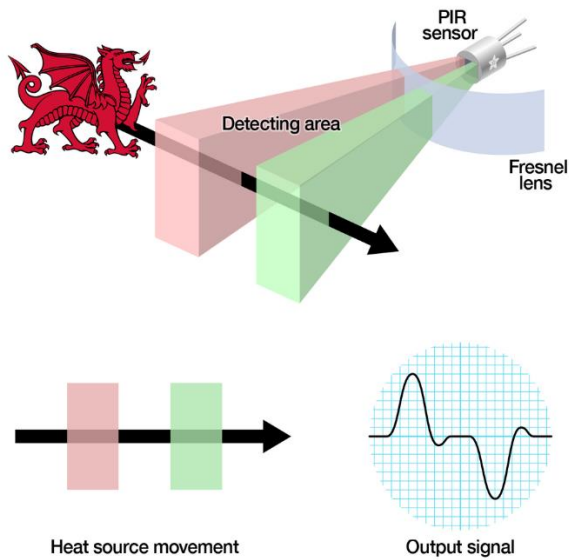

Figure 6: Working of PIR Motion Sensor

However, as an object such as a human or an animal approaches the sensor, the object intercepts one half of the PIR sensor, which causes a positive differential change between the two halves. As the object leaves the sensing are, the opposite happens and the sensor generates a negative differential change. The positive and negative differential changes trigger an output signal as show in figure 5. As a result of the pulse changes, the PIR Sensor triggers an output pulse which is typically 3V in magnitude. In addition to the sensors, the lens on the PIR sensors is another critical component which aids in effective function of the device. The lens around the sensors is a Fresnel lens. The use of Fresnel lens reduces the amount of material required compared to a conventional lens by dividing the lens into a set of concentric annular sections.

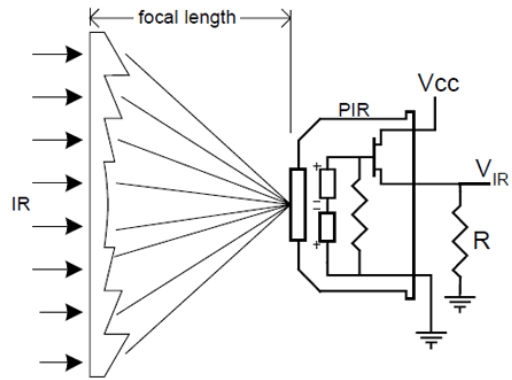

Figure 7: Fresnel Lens

Fresnel lenses have the ability to condense light and provide a larger area of IR for the sensor to detect.

### 2.2.3 Typical Specifications of PIR Sensors

- **Operating voltage:** DC 5 - 12V
- **Static power consumption:** 65 mA
- **Output Signal:** 3V TTL
- **Detection Distance:** Up to 6 meter (Adjustable)
- **Sensing Range:** Less than 120° degree angle, 6 meters
- **Delay Time:** 5 - 200s (Adjustable)
- **Adjustable Trigger:** L: non repeatable trigger - H: repeatable trigger
- **Operating Temperature:** -15 to +70°C
- **Dimensions:** 32 x 24mm, screw hole distance 28mm

# 3 Project Implementation

## 3.1 Use Cases for the Implementation

The implementation of the project focuses primarily on two use cases:

- Image capturing
- Video recording

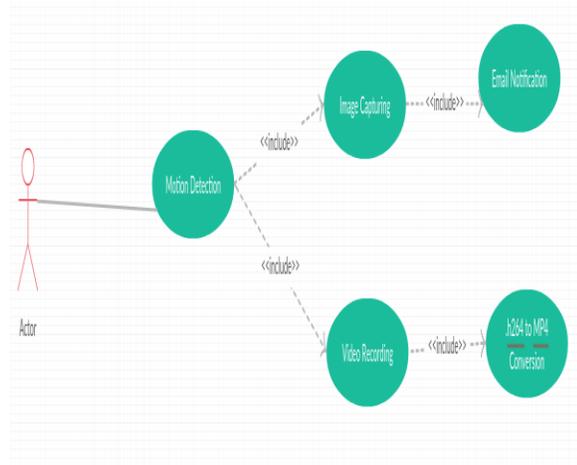

Figure 8: Use Case Diagram Depicting the Application Flows

**Image Capturing:** The focus of the image capturing use case is to detect motion using the PIR motion sensors, take a picture using the Raspberry Pi camera and send the picture to the user as an attachment in an email.

**Video Recording:** The focus of the video recording use case is to detect motion using the PIR motion sensors, record a video for a set amount of time (the project sets the time interval at 5 seconds) and convert the video to MP4 format from h264 format for ease of consumption.

The data flow for the use cases, circuit setup, the software tool setup and the Python code required to achieve the implementation are discussed in the sections below.

## 3.2 Connecting Raspberry Pi to a Shared Network

The following steps should be followed to put the Raspberry Pi on a shared network:

- Connect the Raspberry Pi to a computer that has an internet connection.
- Go to Control Panel -> Network and Internet -> Network Connections and select the network that the computer is connected to.

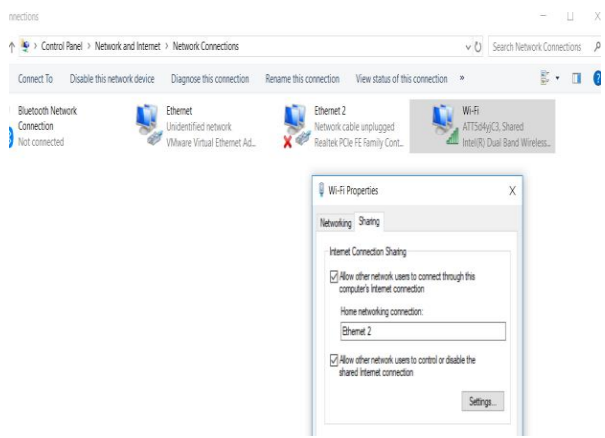

Figure 9: Network Connection Sharing

- Open the network properties and click on Sharing as shown in the picture above.
- Check the checkbox 'Allow other network users to connect through this computer's network connection' and pick a network from the drop-down list.
- Upon doing so, a local area network shows up which is created for the Raspberry Pi.
- The gateway IP of the network should be obtained from the properties of the new network.
- In order to know the IP address assigned to the Pi, a tool called Advanced IP Scanner is used (The

tool can be downloaded from the internet).

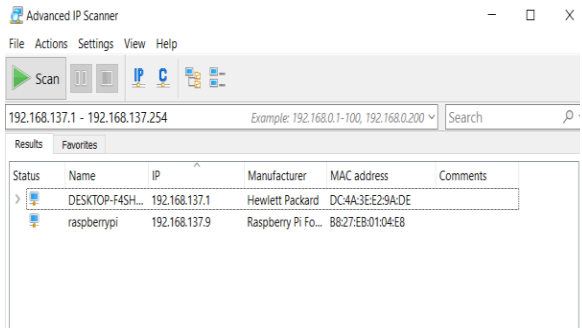

Figure 10: Advanced IP Scanner Showing Raspberry Pi's IP Address

## 3.3 Connecting Raspberry Pi to a Computer Display

In order to connect the Raspberry Pi to a computer display, different tools like Putty, VNC Server or Windows RDP could be used. For this purposes of this project, VNC Server is used. The benefit of using VNC is that it shows a GUI of Raspberry Pi which is easier to work on than Putty which does not provide a GUI.

The following steps are to be followed to install VNC Server onto Raspberry Pi:

- Connect to the Raspberry Pi using the IP address of the Raspberry Pi obtained from AdvancedIPScanner.
- On the Putty terminal, login to the pi using the Pi's credentials and then install VNC Server using the following command:
  **sudo apt-get install tightvncserver**
- Once VNC Server is installed, it can be started using the command:
  **tightvncserver**

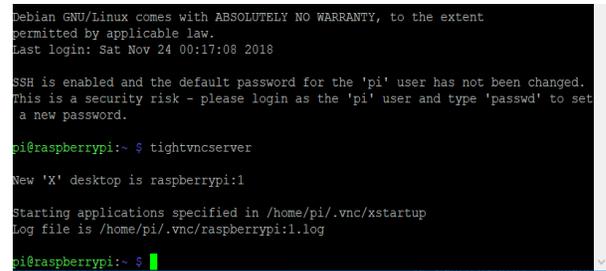

Figure 11: Start Up of VNC Server on Raspberry Pi

- Once the VNC Server is started up, the next step is to remote into the Raspberry Pi using the VNC Viewer. VNC Viewer takes in the IP address of the Raspberry Pi and remotes into the Pi. The GUI of the Raspberry Pi can then be viewed on the computer display.

## 3.4 Lighting an LED Using Raspberry Pi

Lighting up an LED using a Raspberry Pi is fairly simple and is one of the most basic tasks that can be performed using the Raspberry Pi. In order to light up an LED the connections between the Raspberry Pi and the LED would look similar to the connections shown in Figure 12.

The connection is established between the Raspberry Pi and the LED using a resistor (typically 550 ohm) to regulate the voltage that passes into the LED from the Raspberry Pi. The Raspberry Pi's GPIO are used to pass the voltage to the LED.

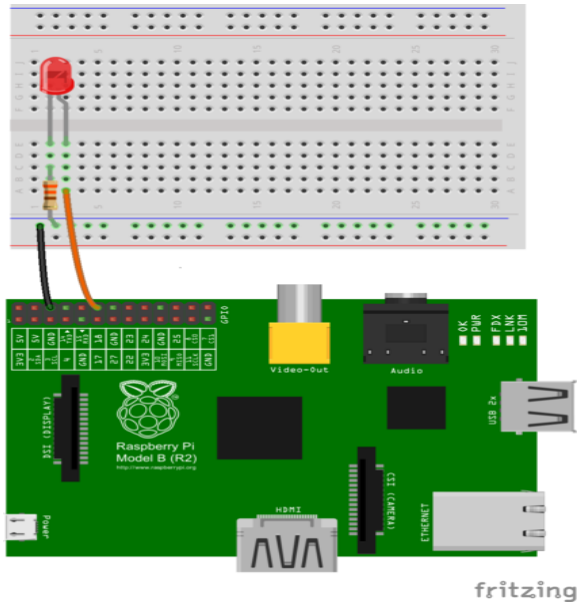

Figure 12: An LED connected to a Raspberry Pi using a breadboard

Once the LED is connected to the Raspberry Pi, a Python program can be used to turn on/off the LED.

The code for the Python program is as follows:

*Import RPi.GPIO as GPIO*

*Import time*

*GPIO.setmode(GPIO.BCM)*

*GPIO.setwarnings(False)*

*GPIO.setup(18, GPIO.OUT)*

*GPIO.input(18, GPIO.HIGH)*

*time.sleep(5)*

*GPIO.output(18, GPIO.LOW)*

The code above turns the LED on for a time interval of 5 seconds and turns the LED back off.

*Note: BCM mode is used to perform the operations on the LED.*

### 3.5 Image Capturing Use Case Implementation

As outlined in section 3.1, the image capturing use case detects motion, takes a picture and sends an email to the user with the picture as an attachment in a nutshell.

The dataflow for the use case is shown in Figure 13.

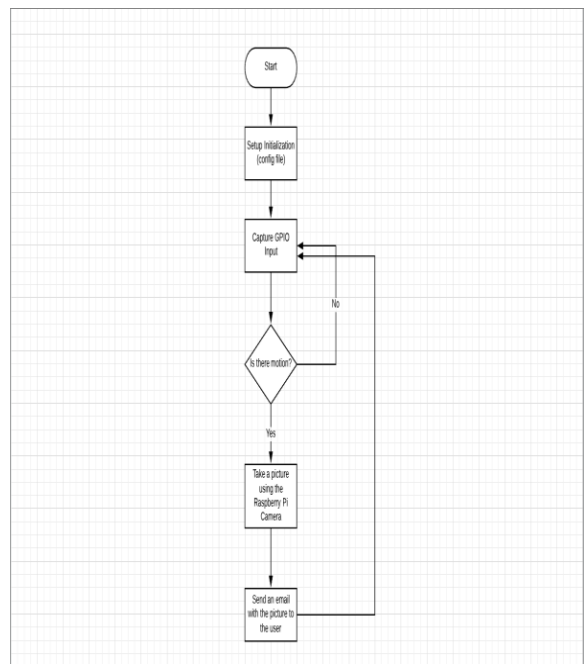

Figure 13: Dataflow Diagram for Image Capture

The sequence of steps to achieve the functionality is:

1. Setup the configuration file for the Python application to read.
2. Write code in Python that performs the following:
   - Parse configuration file setup in step 1.
   - Detects motion using the PIR motion sensor

- Captures an image using the Raspberry Pi camera
- Sends an email with the captured image added as an attachment.

### 3.5.1 Configuration File Setup

The configuration file setup can be done by using a tool named ConfigParser.

**Note: ConfigParser is built into Python3 and does not require installation. However, older versions of Python may or may not have it built in and may require manual installation.**

The setup of the configuration can be done programmatically using ConfigParser or manually following the file format for .ini files which Python can interpret using ConfigParser.

For the purposes of this project, ConfigParser has been used to create the configuration file.

### 3.5.2 Configuration File Parsing

The configuration file that is setup can be parsed using the ConfigParser tool and the properties setup in the configuration file could be used as parameters for the application.

**Note: While the application could be built without using configuration files, the primary reason for using configuration files in this implementation is that Raspberry Pi's email functionality requires the user to specify the email ID and password in plain text. In order to add an extra layer of protection, a configuration file is used so the user credentials can be stored in a file that is not within the code.**

*#Creates a configparser instance and reads the config file config.ini*

*parser = ConfigParser()*

*parser.read('config.ini')*

*#Reads the properties for the email addresses and pasword from the config file*

*email_sender = parser.get('Motion', 'email_sender')*

*email_recipient = parser.get('Motion', 'email_recipient')*

*password = parser.get('Motion', 'password')*

### 3.5.3 Motion Detection Using PIR Motion Sensors

The architecture of the PIR motion sensor as outlined in section 2.2 is leveraged to establish the connection between the Raspberry Pi and the PIR motion sensor.

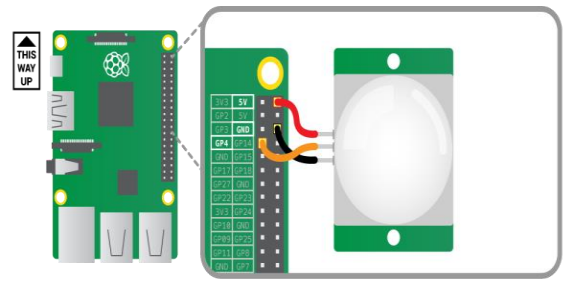

Figure 14: PIR Motion Sensor Connected to Raspberry Pi

The connection between the PIR motion sensor and the Raspberry Pi is shown in Figure 14. Drawing inferences from the figure, the setup involves connecting the voltage pin on the PIR motion sensor to one of the DC voltage pins on the Raspberry Pi – Red jumper cable, the GND pin on the sensor to one of the GND pins on the Raspberry Pi – Black jumper cable and the Digital OUT pin on the PIR motion sensor to one of the GPIO input pins on the Raspberry Pi – Yellow jumper cable.

*#Captures the input on the GPIO pin 11*

*i = GPIO.input(11)*

*#An input of 0 on the GPIO pin 11 indicates there is no motion*

*if i == 0:*

   *print("No Motion Detected")*

   *time.sleep(1)*

*#An input of 1 on the GPIO pin 11 indicates motion is detected*

*elif i == 1:*

   *print("Motion Detected")*

### 3.5.4 Capture Images upon Motion Detection

Upon the detection of motion by the PIR motion sensor, the Raspberry Pi's camera is used to capture an image.

The connection between the Raspberry Pi's camera and the Raspberry Pi is straightforward. The camera strip's Blue color side should be towards the Raspberry Pi's Ethernet port.

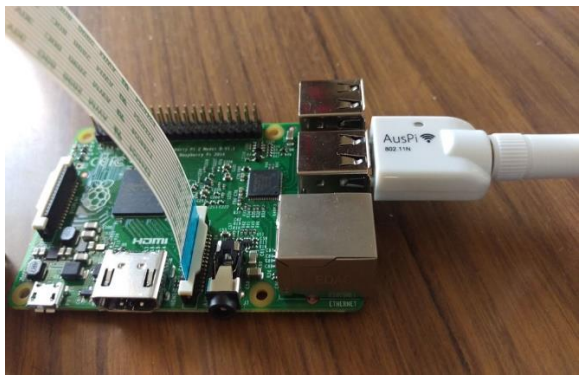

Figure 15: Raspberry Pi's camera connected to Raspberry Pi

The use of the Raspberry Pi camera requires the camera to be enabled on the Raspberry Pi.

In order to enable the camera on the Raspberry Pi, the following steps should be followed:

- Execute the command: **$sudo raspi-config** to open up the configuration of the Raspberry Pi.
- From the options listed in the configuration, select the camera option and enable it as shown in the diagram below.

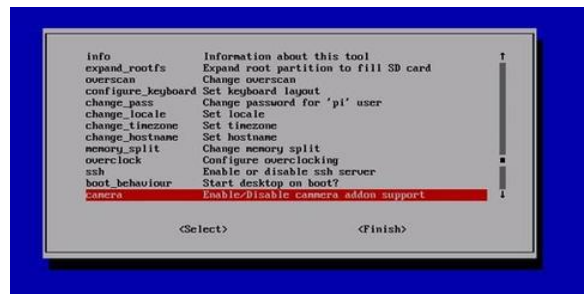

Figure 16: Enabling the Raspberry Pi's Camera

*#Setup the Raspberry Pi camera instance and takes a picture with a resolution of 1280 * 720*

   *with picamera.PiCamera() as camera:*

      *print("About to take a picture")*

      *camera.resolution = (1280, 720)*

      *camera.vflip = True*

*camera.capture("MotionDetected.png")*

   *print("Picture taken")*

Capturing the image requires the resolution to be set (HD resolution of 1280 * 720 is used in the implementation of this project). In addition to the resolution, a property named VFlip is set to True. This is done to ensure that the image that is captured faces upwards rather than downwards. Images captured using the Raspberry Pi camera typically face downwards.

### 3.5.5 Sending Emails with the Image Captured as an Attachment

The image that is captured upon motion detection is then sent over to an email ID of choice. In order to send an email using Raspberry Pi, SMTPLib is used.

**Note: SMTPLib is built-in to Python3 and can be used by importing into the Python application.**

The following code can be used to send an email with an attachment using Python:

```
#Function that sends an email with an attachment
def SendEmail():
    msg = MIMEMultipart()
    msg['From'] = email_sender
    msg['To'] = email_recipient
    msg['Subject'] = subject

    body = 'This is an email from the motion detector app.'
    msg.attach(MIMEText(body, 'plain'))

    filename = 'MotionDetected.png'
    attachment = open(filename, 'rb')

    part = MIMEBase('application', 'octet-stream')
    part.set_payload((attachment).read())
    encoders.encode_base64(part)
    part.add_header('Content-Disposition', "attachment; filename = " + filename)

    msg.attach(part)

    text = msg.as_string()
    server = smtplib.SMTP('smtp.gmail.com', 587)

    server.starttls()
    server.login(email_sender, password)
    server.sendmail(email_sender, email_recipient, text)
    server.quit()
```

While sending an email using Python is a relatively easy task, there are two caveats to it. They are:

- Python requires that the user credentials for the email account be specified in plain text.
- Any email that is required to be sent to a Gmail ID needs the 'Access for less secure apps' option to be turned ON. This results in a lowered amount of security.

For the purposes of this project, the user credentials are read from a configuration file using the tool, ConfigParser. This approach meant that the credentials did not have to be present in code making the code cleaner.

### 3.6 Video Recording Use Case Implementation

As outlined in section 3.1, the goal of the use case is to record video for a specified interval of time upon motion detection and to convert the video to MP4 format for ease of use.

The reason for the conversion is that Raspberry Pi natively records video in .h264 format which is not supported by most video players such as Windows Media Player on Windows or QuickTime on Mac. The conversion into MP4 format means that these media players can play the video.

The dataflow for the use case is as shown in figure 17.

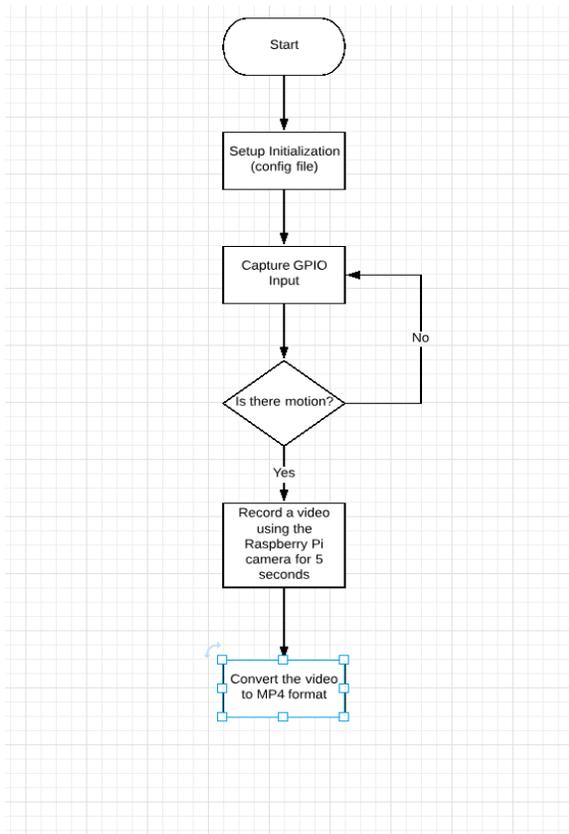

Figure 17: Dataflow Diagram for Video Recording

When we compare the video recording dataflow with the image capture dataflow, it can be seen that both of the use cases have the same dataflow up till the point of motion detection. Hence, the code up till motion detection as described in the Image Capture use case can be used in this use case as well.

### 3.6.1 Video Recording upon Motion Detection

Recording a video upon motion detection is achieved by a Python program. The code shown below records a video for a time period of 5 seconds (the time period is customizable).

*print("About to record video")*

    *camera.resolution = (1280, 720)*

    *camera.vflip = True*

*camera.start_recording("/home/pi/Videos/motiondetection.h264")*

    *time.sleep(5)*

    *camera.stop_recording()*

    *print("Video recorded")*

### 3.6.2 Conversion from .h264 Format to MP4 Format

The video that is recorded by Raspberry Pi is natively in .h264 format. In order to convert the video from .h264 format to MP4 format, a tool named GPAC is used. GPAC is not built-in and needs added to the Raspberry Pi.

GPAC can be added to the Raspberry Pi using the following command:

**sudo apt-get install gpac**

The code below can be used to record video using the Raspberry Pi. For the purposes of this project, a 5 second recording time frame is used.

*#Converting the video to MP4*

    *print("Converting video to MP4")*

    *command = "MP4Box -add /home/pi/Videos/motiondetection.h264 /home/pi/Videos/MotionDetectionConverted.mp4"*

    *call([command], shell = True)*

    *print("Video converted to MP4")*

## 4 UPPAAL Model Verification

The use of UPPAAL is employed to verify that the dataflow of the system that is built is indeed correct and that the system works as expected. The generated model for the system is shown below.

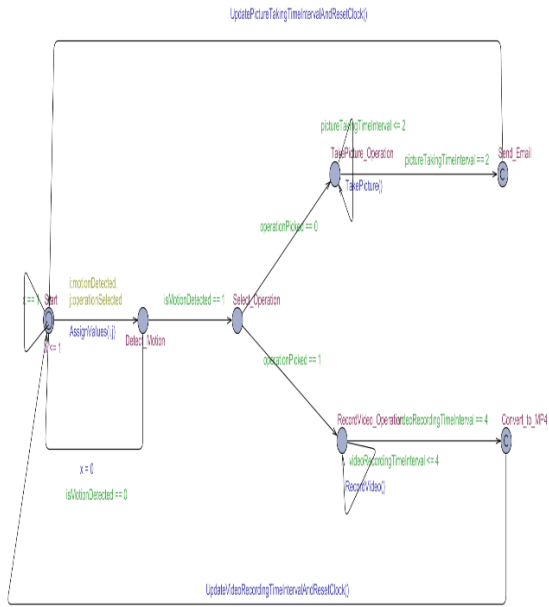

Figure 18: UPPAAL Model Verifying the System

The UPPAAL model shown above randomly picks a value that is used to determine if there is motion detected by the system or not. As and when the model picks the value that indicates that there is motion (0 is used to indicate there is no motion and 1 is used to indicate that there is motion), the model then checks to see which operation needs to be executed where operation refers to a use case (the system has two use cases – picture capturing and video recording). Based on the use case that is selected, the model then executes the corresponding use case and then sends the control back to the start state where the system starts to check to see if there is motion detected and the cycle begins once again. Note that the Send_Email state on the picture capturing use case and the Convert_to_MP4 state on the video recording use case are committed states to ensure that they leave the states immediately since those operations don't take a significant amount of time.

The model is verified using the following queries which show that the model behaves as expected and described in the previous paragraph.

```
Overview
E<>(operationPicked == 0)
E<>(operationPicked == 1)
E[](pictureTakingTimeInterval <= 2)
E[](videoRecordingTimeInterval <= 4 )
A[](not deadlock)

Query
A[](not deadlock)
```

Figure 19: Queries to Verify the Model

The E[](videoRecodringTimeInterval <= 4) query in the above picture shows that the model records videos for a time interval of 5 time units on all paths eventually and the E[](pictureTakingTimeInterval <= 2) query shows that the picture capturing takes a time interval of 2 time units on all paths eventually.

## 5 Next Steps

Going forward, there is scope for improvement in the solution. At least three steps for improvement have been identified and are listed as follows:

- Development of an encryption mechanism to ensure that user credentials are not sent in plain text.
- Extending the notification capability of the application to text messaging.
- Enhancing the video recording feature to record audio input.

## 6 Conclusion

Smart security systems aid in providing better awareness about the surroundings and help people protect themselves from potential hazards in many different ways. The solution described in this paper is effective while also being low cost and low maintenance. The use of Raspberry Pi also has the benefits of high scalability and the availability of a plethora of built-in components that could be integrated for further enhancements to the project.